# Effect of Heat Treatment Mode and Aggressive Media on Mechanical Properties of Porous Polytetrafluoroethylene Membranes Fabricated via Electrospinning

*Evgeniy Melnik[1], Ksenia Stankevich[2], Aleksey Zinoviev[1], Ekaterina Poletykina[1], Artem Andreev[1], Vyacheslav Bouznik[3], Evgeniy Bolbasov[1,4]\**

1. Tomsk Polytechnic University, 634050, Tomsk, Russian Federation.
2. Montana State University, Bozeman, MT 59717, USA
3. Tomsk State University, 634050 Tomsk, Russian Federation.
4. V.E. Zuev Institute of Atmospheric Optics SB RAS, 634055, Tomsk, Russian Federation.

\* Corresponding author: Evgeniy Bolbasov. Email: Ftoroplast@tpu.ru

**Abstract**

Electrospinning is a modern alternative to the expanded method for producing porous polytetrafluoroethylene membranes. High strength and relative elongation, as well as the ability to maintain these properties for a long time when exposed to aggressive media at high temperatures, determine the application scope of the electrospun polytetrafluoroethylene membranes. Herein, we report the effect of polytetrafluoroethylene suspension content in the spinning solution, heat treatment mode (quenching and annealing) and aggressive media at high temperatures on the tensile strength and relative elongation of electrospun polytetrafluoroethylene membranes. Membranes fabricated from spinning solutions with 50 to 60 wt % polytetrafluoroethylene suspension content that underwent quenching were characterized by the highest tensile strength and relative elongation. Electrospun polytetrafluoroethylene membranes also demonstrated high chemical resistance to concentrated mineral acids and alkalis, a bipolar aprotic solvent, engine oil and deionized water at 100 °C for 48 hours.

**1. Introduction.**

Due to high chemical resistance, thermal stability, excellent dielectric properties and biocompatibility, porous polytetrafluoroethylene (PTFE) membranes are widely used in the chemical industry, hydrogen and nuclear energy, aviation and space industries, reconstructive surgery, etc. [1]. Although expanded remained one of the most effective methods for producing PTFE membranes with the desired technical characteristics [2], electrospinning is becoming a popular alternative. This is a two-step technique that includes fabrication of a composite membrane from a polymer with good fiber-forming properties (e.g., polyvinyl alcohol (PVA) or high molecular weight polyethylene oxide (PEO)) and PTFE particles (commonly, water suspensions of PTFE) that is then subjected to the temperatures higher than 330 °C. During the heat exposure, PVA/PEO are removed from the membrane resulting in sintering of PTFE particles with the formation of PTFE fibers. Practical potential of electrospun PTFE membranes is being currently explored including their application for membrane distillation process [3], membrane emulsification [4], self-cleaning surfaces [5], as a separator for lithium-ion batteries [6], and the use in biomaterial area [7].

Strength and relative elongation are the key properties determining application scope of electrospun PTFE membranes. It is also important that these properties would be retained when membranes are exposed to aggressive media at high temperatures for a long time.

Since PTFE is a semi-crystalline polymer, the mechanical properties of PTFE-based materials are determined by its degree of crystallinity, which, in turn, is dependent on the cooling rate during the heat treatment [8]. Thus, the cooling rate of PTFE membranes formed during the heat treatment of electrospun PTFE/PVA(PEO) membranes should affect the mechanical properties of the former.

The influence of cooling rate on the strength and relative elongation of electrospun PTFE membranes obtained from the composite membranes with various component ratios has not been systematically studied. The available data provide limited insight on the effect of the exposure to aggressive media (NaOH (6 mol $L^{-1}$) and $H_2SO_4$ (7.14 mol $L^{-1}$) at 100 °C for 24 hours) on the mechanical properties of PTFE membranes produced by the heat treatment of PTFE/PEO electrospun membranes with a certain PTFE/PEO ratio [9].

The lack of comprehensive studies that consider the impact of both cooling rate and aggressive media on mechanical properties of PTFE membranes limit their practical application. Herein, we report the results of a systematic study of the effect quenching and annealing and exposure to aggressive media at high temperatures on the tensile strength and relative elongation of electrospun PTFE membranes obtained by the heat treatment of PTFE/PVA membranes with various PTFE/PVA ratios.

**2. Material and methods**

Spinning solutions were made by mixing a 10% aqueous solution of polyvinyl alcohol (PVA) (JSC Vekton, Russia) with an aqueous suspension of PTFE (Halopolymer, Russia) as described previously [7]. The following PVA/PTFE ratios were used: 50/50, 40/60, 30/70 and 20/80 wt %.

PTFE membranes were obtained in a two-step procedure. First step - fabrication of the composite PVA/PTFE membranes using NANON-01A electrospinning system (MECC CO, Japan). Electrospinning parameters: distance from the injector (needle 22 G) to the cylindrical collector (diameter - 100 mm, width - 200 mm) - 16 cm, collector rotation speed - 200 rpm, spinning solution flow rate - 1.2 mL/hour, voltage - 27 kV.

Second step included heat treatment of the composite membranes under two sets of conditions referred to as quenching and annealing. During quenching composite membranes were heated over 3 hours to 330 °C, kept at 330 °C for 10 minutes, then removed from the oven and cooled to a room temperature. During annealing composite membranes were heated over 3 hours to 330 °C, and then cooled to a room temperature inside the oven over 14 hours.

To eliminate interference of PVA degradation products with the study results, all PTFE membranes were placed in perfluorodecalin (Ekros 1, Russia) and treated with gaseous mixture of nitrogen and fluorine (fluorine concentration 10% vol.) for 4 hours at room temperature. After treatment the membranes were washed with ethanol and placed in a vacuum oven (Aktan Vacuum, Russia), where they were kept at 50°C and 15 kPa for 12 hours.

The impact of the following aggressive media on PTFE membranes was studied: nitric acid ($HNO_3$ 10.3 mol $L^{-1}$), potassium hydroxide (KOH 11.6 mol $L^{-1}$), dimethyl sulfoxide (DMSO), motor oil and deionized water ($H_2O$). During the experiment, PTFE membranes were kept in the medium at 100 °C for 48 hours.

The effect of heat treatment and exposure to aggressive media on the tensile strength and relative elongation of PTFE membranes was evaluated using an Instron 3369 testing machine (Instron, USA) equipped with a load cell # 2519-102 (Instron, USA) at a sample tension rate of 10 mm/min. Samples were obtained using a ZCP 020 punching press (ZwickRoell, Germany). Each sample group contained seven repeats.

Statistical analysis of the obtained results was carried out using the Kruskal-Wallis test in the GraphPad Prism 8 software.

**2. Results and discussion**

Table 1 summarizes the effect of the heat treatment mode and spinning solution composition on the strength and relative elongation of PTFE membranes. The study shows that an increase in the PTFE suspension content in the spinning solution leads to a decrease in tensile strength and relative elongation of the formed PTFE membranes regardless of the heat treatment mode. The obtained results can be rationalized by an increased mobility of PTFE macromolecules on the surface of PTFE suspension particles due to the plasticizing effect of PVA degradation products released during the heat treatment of the PTFE/PVA membrane. The plasticizing effect facilitates sintering of PTFE particles, contributing to the formation of fibers with fewer defects [7].

Table 1. Tensile strength and relative elongation of electrospun PTFE membranes depending on PTFE suspension content and heat treatment mode.

| PTFE suspension content, wt % | Tensile strength, MPa | Elongation at break, % |
|---|---|---|
| Quenching | | |
| 50 | 2.32±0.08 | 324±39 |
| 60 | 2.14±0.17 | 309±37 |
| 70 | 1.90±0.14 | 268±52 |
| 80 | 1.65±0.13 | 226±16 |
| Annealing | | |
| 50 | 1.35±0.32 | 142±10 |
| 60 | 1.12±0.24 | 83±15 |
| 70 | 0.91±0.21 | 90±22 |
| 80 | 0.72±0.12 | 76±16 |

While the similar trend in mechanical properties versus PTFE suspension content is observed for membranes that underwent both quenching and annealing, the former exhibit higher tensile strength and relative elongation comparing to the latter (Table 1). This can be explained by the effect of the cooling rate on PTFE crystallinity degree during the heat treatment. It is known that the highest crystallization rate of PTFE is observed at ~300 ÷ 310 °C [8]. Since during crystallization membranes were cooled slowly, they would have a higher crystallinity degree and a more perfect crystal structure compared to membranes obtained after quenching. Crystallization is accompanied by a significant shrinkage and an increase in polymer density leading to an increase in internal stress in the fibers that form the membrane. This, in turn, results in the formation of a significant number of defects reducing the tensile strength of crystallized PTFE membranes. It should be noted that the quenching membranes obtained from spinning solutions containing from 50 to 60 wt % of PTFE suspension had the best mechanical properties from all the studied groups (Table 1).

We further investigated the effect of aggressive media at high temperatures on the tensile strength and relative elongation of PTFE membranes obtained during quenching as they showed better mechanical properties. During this study membranes were kept in various aggressive media for 48 hours at 100 °C and then mechanical properties were evaluated. Table 2 summarizes the obtained results.

Table 2. Tensile strength and relative elongation of electrospun PTFE membranes after exposure to aggressive media.

| PTFE suspension content, wt % | Control | HNO$_3$ | DMSO | Motor oil | KOH | H$_2$O |
|---|---|---|---|---|---|---|
| Elongation at break, % | | | | | | |
| PTFE 50 | 334±39 | 355±22 | 332 ±24 | 361±42 | 312±29 | 304±22 |
| PTFE 60 | 298±31 | 300±36 | 293±49 | 319±42 | 315±21 | 322±26 |
| PTFE 70 | 260±21 | 278±28 | 264 ±18 | 241±45 | 255±27 | 247±35 |
| PTFE 80 | 216±26 | 23 ±34 | 222±13 | 208±28 | 212±26 | 230±27 |
| Tensile strength, MPa | | | | | | |
| PTFE 50 | 2.26±0.18 | 2.38±0.41 | 2.18±0.37 | 2.27±0.32 | 2.17±0.23 | 2.28±0.31 |
| PTFE 60 | 2.00±0.33 | 2.09±0.39 | 2.14±0.32 | 2.07±0.18 | 1.95±0.24 | 2.12±0.27 |
| PTFE 70 | 1.85±0.26 | 1.73±0.18 | 1.92±0.25 | 1.94±0.21 | 2.00±0.19 | 1.91±0.31 |
| PTFE 80 | 1.53±0.31 | 1.55±0.23 | 1.60±0.31 | 1.55±0.30 | 1.63±0.35 | 1.61±0.31 |

According to the data, PTFE membranes retained their mechanical properties regardless of PTFE suspension content in the spinning solution and the aggressive medium used (Table 2). The absence of significant changes in strength and relative elongation of electrospun PTFE membranes is due to the following. As PVA thermal degradation products were completely removed from the membranes during the treatment with a gaseous mixture of nitrogen and fluorine, aggressive medium interacted with pure PTFE, which is known for its chemical resistance. Since no polymer destruction occurred during the exposure to the aggressive medium, membranes preserved their mechanical properties.

### 3. Conclusions

In summary, we studied the effect of polytetrafluoroethylene suspension content in the spinning solution, heat exposure (quenching and annealing) and aggressive media at high temperatures on the tensile strength and relative elongation of electrospun polytetrafluoroethylene membranes. We found that decreased content of polytetrafluoroethylene suspension in the spinning solution and quenching improve tensile strength and relative elongation of polytetrafluoroethylene membranes. Membranes fabricated from spinning solutions with 50 to 60 wt % polytetrafluoroethylene suspension content that underwent quenching had the best mechanical properties among the groups studied. Exposure of electrospun polytetrafluoroethylene membranes to concentrated mineral acids and alkalis, a bipolar aprotic solvent, engine oil and deionized water at 100 °C for 48 hours did not deteriorate their tensile strength and relative elongation, demonstrating their high chemical resistance.

**Acknowledgments**

The reported study was funded by RFBR, project number 20-03-00171.

**References**

[1] S. Feng at al. Progress and perspectives in PTFE membrane: Preparation, modification, and applications, J. Memb. Sci. 549 (2018) 332–349. https://doi.org/10.1016/j.memsci.2017.12.032.

[2] S. Ebnesajjad, History of Expanded Polytetrafluoroethylene and W.L. Gore & Associates, in: Introd. to Fluoropolymers, Elsevier, 2021: pp. 33–42. https://doi.org/10.1016/B978-0-12-819123-1.00004-5.

[3] C. Su at al. Novel PTFE hollow fiber membrane fabricated by emulsion electrospinning and sintering for membrane distillation, J. Memb. Sci. 583 (2019) 200–208. https://doi.org/10.1016/j.memsci.2019.04.037.

[4] S. Yu at al. Pore structure optimization of electrospun PTFE nanofiber membrane and its application in membrane emulsification, Sep. Purif. Technol. 251 (2020) 117297. https://doi.org/10.1016/j.seppur.2020.117297.

[5] F. Zou at al., Dynamic hydrophobicity of superhydrophobic PTFE-SiO2 electrospun fibrous membranes, J. Memb. Sci. 619 (2021) 118810. https://doi.org/10.1016/j.memsci.2020.118810.

[6] J. Li at al. Electrochemical performance and thermal stability of the electrospun PTFE nanofiber separator for lithium-ion batteries, J. Appl. Polym. Sci. 135 (2018) 46508. https://doi.org/10.1002/app.46508.

[7] I. Kolesnik at al. Characterization and Determination of the Biocompatibility of Porous Polytetrafluoroethylene Membranes Fabricated via Electrospinning, J. Fluor. Chem. 246 (2021) 109798. https://doi.org/10.1016/j.jfluchem.2021.109798.

[8] G.J. Puts at al. Polytetrafluoroethylene: Synthesis and Characterization of the Original Extreme Polymer, Chem. Rev. 119 (2019) 1763–1805. https://doi.org/10.1021/acs.chemrev.8b00458.

[9] Y. Feng at al. Mechanical properties and chemical resistance of electrospun polyterafluoroethylene fibres, RSC Adv. 6 (2016) 24250–24256. https://doi.org/10.1039/C5RA27676D.